# Stochastic resonance of a nanomagnet excited by spin transfer torque

XIAO CHENG, CARL T. BOONE, JIAN ZHU, AND ILYA N. KRIVOROTOV[†]

Department of Physics & Astronomy, University of California, Irvine, California 92617, USA

[†] ikrivoro@uci.edu

**Spin transfer torque from spin-polarized electrical current can excite large-amplitude magnetization dynamics in metallic ferromagnets of nanoscale dimensions. Since magnetic anisotropy energies of nanomagnets are comparable to the thermal energy scale, temperature can have a profound effect on the dynamics of a nanomagnet driven by spin transfer torque. Here we report the observation of unusual types of microwave-frequency nonlinear magnetization dynamics co-excited by alternating spin transfer torque and thermal fluctuations. In these dynamics, temperature amplifies the amplitude of GHz-range precession of magnetization and enables excitation of highly nonlinear dynamical states of magnetization by weak alternating spin transfer torque. We explain these thermally activated dynamics in terms of non-adiabatic stochastic resonance of magnetization driven by spin transfer torque. This type of magnetic stochastic resonance may find use in sensitive nanometer-scale microwave signal detectors.**

Spin transfer torque (STT) from spin-polarized current[1,2,3,4] applied to a ferromagnet can give rise to several types of magnetization dynamics such as auto-oscillations of magnetization excited by direct current[5,6,7,8,9,10,11], magnetization reversal driven by STT[12,13,14,15,16,17] and current-induced magnetic domain wall motion[18,19,20]. Practical applications of STT in non-volatile memory, microwave oscillators and detectors are under development[21], and non-equilibrium states of magnetization induced by STT are of fundamental interest in nonlinear science[22,23,24,25]. Alternating STT is employed in spin torque ferromagnetic resonance (st-FMR)[26,27,28] measurements of spin waves in magnetic nanostructures[26,29]. In this paper, we describe unusual nonlinear GHz-range magnetization dynamics that are co-excited by ac STT and temperature. The pronounced thermally activated character sets these dynamics apart from the types of excitations seen in st-FMR and conventional FMR experiments. Temperature-dependent studies reveal that these dynamics arise from non-adiabatic stochastic resonance (NSR) of magnetization excited by ac STT. The amplitude of the observed NSR exceeds that of typical st-FMR-driven excitations by nearly two orders of magnitude for the same level of ac STT drive.

We study magnetization dynamics of the permalloy (Py=$Ni_{81}Fe_{19}$) free layer in Pt(30 nm)/ Cu(20 nm)/ Py(3 nm)/ Cu(6 nm)/ Co(3 nm)/ $Ir_{20}Mn_{80}$(8 nm)/ Cu(80 nm) spin valve (SV) nanopillars. All magnetic layers are patterned into 120×60 $nm^2$ elliptical pillars with the long axis parallel to the exchange bias field from $Ir_{20}Mn_{80}$. Electrical current, $I$, is applied between the top and the grounded bottom leads as shown in Fig. 1a. We report data from a single sample; nine other samples show similar behavior. For all measurements, a magnetic field, $\vec{H}$, is applied at 10° from the sample normal, with the in-plane field component in the exchange bias direction. Fig. 1b shows resistance, $R$, of the SV as a function of $H$ at temperatures of 295 K and 120 K. The in-plane component of $\vec{H}$ switches the magnetization of the uniaxial Py nanomagnet, $\vec{M}$, between the high-resistance, nearly antiparallel (HR), and the low-resistance, nearly parallel (LR), states of the SV. The LR↔HR transition is continuous at 295 K and hysteretic at 120 K, indicating that the free layer is superparamagnetic at 295 K. Figures 1c and 1d illustrate the LR↔HR transition induced by STT[1], $\vec{\tau} \sim I_{dc} \cdot \vec{M} \times (\vec{M} \times \vec{M}_{Co})$, from direct current, $I_{dc}$. Fig. 1e shows differential resistance, $dV/dI$, versus $H$ at 120 K measured for several $I_{dc}$. This figure illustrates that the free layer hysteresis loop width decreases with increasing $|I_{dc}|$ and becomes zero at $I_{dc}$ = -2.6 mA, indicating that the free layer is superparamagnetic for $I_{dc} \leq$ -2.6 mA. Figure 1f shows telegraph noise of the free layer between the HR and LR states at $T$ = 120 K, $H$ = 2.2 kOe and $I_{dc}$ = -2.7 mA, which demonstrates that the free layer indeed becomes superparamagnetic at this current. This current-induced superparamagnetism results from renormalization of the amplitude of thermal fluctuations of $\vec{M}$ by STT[14,30]. For $I_{dc} <$ 0, STT pushes $\vec{M}$ away from the LR state, thereby enhancing the amplitude of thermal fluctuations in this state, and increases the rate of thermally activated transitions out of the LR state. In contrast, the amplitude of thermal fluctuations in the HR state is suppressed for $I_{dc} <$ 0 because STT pushes magnetization towards the HR state and thereby decreases the rate of thermally activated transitions out of the HR state. As a result, the Kramers transition rates between the HR and LR states become functions of current[30]:

$$K_{HR \to LR} \approx f_0 \exp\left(-\left(1 - \frac{I}{|I_0^{HR}|}\right)\frac{E_b^{HR}(H)}{k_B T}\right) \quad (1a),$$

$$K_{LR \to HR} \approx f_0 \exp\left(-\left(1 + \frac{I}{|I_0^{LR}|}\right)\frac{E_b^{LR}(H)}{k_B T}\right) \quad (1b),$$

where $E_b^{HR(LR)}(H)$ are the zero-current barriers for transition out of the HR(LR) states, $k_B$ is the Boltzmann constant, $f_0 \approx 10^9$-$10^{10}$ Hz and $I_0^{HR(LR)} = const$ are the critical currents[30].

We measure magnetization dynamics driven by ac STT using the setup in Fig. 2a[26]. In these measurements, ac STT from a square-wave modulated microwave current of rms amplitude $I_{ac}$ is applied to the SV in addition to $I_{dc}$. The ac STT excites oscillations of $\vec{M}$, which give rise to oscillations of $R$ via giant magnetoresistance, $R = R_0 + \delta R_{ac} \sin(\omega t + \varphi)$, where $\varphi$ is the phase shift between the resistance and current oscillations. As a result, a rectified voltage $\frac{1}{\sqrt{2}} I_{ac} \delta R_{ac} \cos(\varphi)$ is generated by the SV. At $I_{dc} \neq$ 0, an additional voltage, $I_{dc} \cdot \delta R_{dc}$, is induced in response to $I_{ac}$, where $\delta R_{dc}$ is the difference between the time-average SV resistance in the presence and in the absence of $I_{ac}$[31]. The resulting rms voltage measured by the lock-in amplifier is $V_{dc} = \frac{1}{\pi} I_{ac} \delta R_{ac} \cos(\varphi) + \frac{\sqrt{2}}{\pi} I_{dc} \delta R_{dc}$, where the first term is dominant for small-amplitude oscillations of $\vec{M}$ while the second term dominates for large-amplitude oscillations of $\vec{M}$, provided $|I_{dc}| \gg I_{ac}$[31]. In our measurements, we sweep the frequency of $I_{ac}$, $f$, and record $V_{dc}(f)$. When $f$ coincides with a spin wave eigenfrequency of the Py nanomagnet, $f_n$, the eigenmode is resonantly excited, and a peak or a trough in $V_{dc}(f)$ is observed at $f = f_n$. These peaks and troughs give the st-FMR spectrum of spin wave eigenmodes of the nanomagnet[26,27].

Figure 2c-e shows $V_{dc}(f)$ at $T = 120$ K and $H = 2.2$ kOe for several values of $I_{dc}$: far from (-1.5 mA, -3.5 mA) and at (-2.6 mA to -2.8 mA) the current-driven LR↔HR transition. The st-FMR spectra, $V_{dc}(f)$, far from the LR↔HR transition shown in the insets of Fig. 2c-d exhibit typical spin wave resonances[26,27,32,33]. However, for $I_{dc}$ at the LR↔HR transition, we observe qualitatively different $V_{dc}(f)$. At $I_{dc} = -2.6$ mA (Fig. 2c), $V_{dc}(f)$ develops a negative low-frequency tail $V_{max}^{LF} \equiv |V_{dc}(0)|$ while at $I_{dc} = -2.8$ mA (Fig. 2d), $V_{dc}(f)$ exhibits a positive maximum, $V_{max}^{HF}$, at $f = 2.2$ GHz. Both $V_{max}^{LF}$ and $V_{max}^{HF}$ are two orders of magnitude greater than the amplitudes of the high-frequency resonances, $V_{max} \equiv max\{|V_{dc}(f)|\}_{f>1GHz}$, in the LR and HR states, indicating that large-amplitude dynamics are excited by $I_{ac}$ at the LR↔HR transition. Fig. 2e shows the response curve at $I_{dc} = -2.7$ mA in the crossover regime between the large-amplitude high- and low-frequency resonances. Fig. 2f displays $V_{dc}(t)$ at $I_{dc} = -2.8$ mA and $f = 2.2$ GHz for $I_{ac} = 0$ mA and $I_{ac} = 0.4$ mA. We also make spectral measurements of microwave signal emission by the sample[5,6] at $I_{dc} = -2.8$ mA and $I_{ac} = 0$ mA, and do not observe a detectable signal in the 0.1-10 GHz band. These measurements show that the large-amplitude oscillations of $\vec{M}$ are only excited in the presence of $I_{ac}$.

Figure 3 further illustrates the difference of the dynamic response at the LR↔HR transition from that in the HR and LR states. Fig. 3a-b shows the full width at half maximum and the amplitude, $V_{max}$, of the high frequency ($f > 1$ GHz) spectral peak in $V_{dc}(f)$ versus $I_{dc}$. For $|I_{dc}| < 2.4$ mA, the linewidth of the peak decreases while $V_{max}$ increases with increasing $|I_{dc}|$. This behavior is due to renormalization of the effective damping by dc STT[26,34]. However, for -3.1 mA < $I_{dc}$ < -2.6 mA, both the amplitude and the line width increase dramatically, signaling a transition to a new dynamic regime. Fig. 3c-d shows the dependence of $V_{max}$ on $I_{ac}$ for several values of $I_{dc}$. For $I_{dc}$ far from the LR↔HR transition, $V_{max}(I_{ac})$ is quadratic. In this small-amplitude regime, $\delta R_{dc} \approx 0$[31] and $V_{max} \approx \frac{1}{\pi} I_{ac} \delta R_{ac} \sim I_{ac}^2$. In contrast, at the LR↔HR transition, $V_{max}(I_{ac})$ crosses over from quadratic to linear behavior and eventually saturates at a value close to $\left|\frac{\sqrt{2} I_{dc} \Delta R}{2\pi}\right|$, where $\Delta R = 35$ mΩ is the resistance difference between the HR and LR states at $I_{dc} = -2.8$ mA. This type of response indicates that $I_{ac}$ induces a transition from the HR state with resistance $R_0 + \Delta R$ to a state with time-average resistance of $\approx R_0 + \Delta R/2$. The large-amplitude high- and low-frequency responses of the types shown in Fig. 2c-d are observed in a strip in the ($I_{dc}$,$H$) plane, which coincides with the region of the LR↔HR transitions as illustrated in Fig. 3e.

The origin of the large-amplitude dynamics at the LR↔HR transition is revealed by the temperature dependence of $V_{dc}(f)$. Fig. 3f shows the maximum amplitudes of the high, $\tilde{V}_{max}^{HF} \equiv \max\{V_{dc}(f,H)\}_{f, H}$, and low, $\tilde{V}_{min}^{LF} \equiv \min\{V_{dc}(0,H)\}_{H}$, frequency resonances versus $T$. Below a threshold temperature, $T_{th}(I_{dc})$, $\tilde{V}_{max}^{HF}$ and $\tilde{V}_{min}^{LF}$ are small and $V_{dc}(f)$ shows small-amplitude eigenmode resonances. Above $T_{th}(I_{dc})$, $\tilde{V}_{max}^{HF}$ and $\left|\tilde{V}_{min}^{LF}\right|$ rapidly rise to their maximum values at $T = T_{SR}^{HF(LF)}(I_{dc})$ and the response curves become similar to those in Fig. 2c-d. For $T > T_{SR}(I_{dc})$, $\tilde{V}_{max}^{HF}$ and $\left|\tilde{V}_{min}^{LF}\right|$ slowly decrease. This type of temperature dependence of the amplitude of motion is a salient feature of stochastic resonance (SR)[36].

SR is an effect of noise-induced amplification of the response of a nonlinear system to a weak periodic drive[35,36,37]. To describe SR, we consider the magnetic energy[38] of the Py nanomagnet (Fig. 4):

$$E = 2\pi \vec{M} \cdot \vec{\vec{N}} \cdot \vec{M} - \vec{M} \cdot (\vec{H} + \vec{H}_d), \tag{2}$$

where $\vec{\vec{N}} = \{N_x, N_y, N_z\}$ is the diagonal demagnetization tensor of the free layer[38], $\vec{H}$ is the external field, $\vec{H}_d \sim \vec{M}_{Co}$ is the stray field from the Co layer and $\vec{M}_{Co}$ is the magnetization of Co. Fig. 4b-d shows that $E(\theta, \phi)$ is an asymmetric double-well potential with the wells corresponding to the HR and LR states.

According to Eq. (1), the Kramers transition rates between the HR and LR states depend on STT: $K_{HR \rightarrow LR}$ decreases and $K_{LR \rightarrow HR}$ increases with increasing $|I_{dc}|$. Therefore, equal dwell times in the HR and LR states ($K_{HR \rightarrow LR} = K_{LR \rightarrow HR} \equiv K_E$) can be achieved via tuning of $I_{dc}$ even though the double-well potential is asymmetric. Analysis of Eq. (1) and previous experiments[30] show that $K_{HR \rightarrow LR} = K_{LR \rightarrow HR}$ is observed on a line $H = H_E(I_{dc})$ in the $(I_{dc}, H)$ plane, and $K_E(I_{dc}, H_E(I_{dc}))$ on this line exponentially increases with $|I_{dc}|$. Alternating current, $I_{ac}$, applied in addition to $I_{dc}$ periodically modulates $K_{HR \rightarrow LR}$ and $K_{LR \rightarrow HR}$, and, due to the exponential sensitivity of the transition rates to current, small $I_{ac}$ can induce periodic transitions between the HR and LR states at the frequency of $I_{ac}$. In this case, the LR↔HR transitions are random at $I_{ac} = 0$ and nearly periodic at $I_{ac} \neq 0$. The periodic LR↔HR transitions induced by $I_{ac}$ cease if $T$ becomes too low ($T < T_{th}(I_{dc})$) so that $K_{HR \rightarrow LR}$ or $K_{LR \rightarrow HR}$ becomes small compared to $f$. This temperature-induced amplification of the amplitude of motion of $\vec{M}$ at $T > T_{th}(I_{dc})$ under the action of weak ac drive is SR[39].

Low-frequency (adiabatic) SR driven by ac STT can explain the large low-frequency tail of $V_{dc}(f)$ in Fig. 2c. Indeed, such a tail is observed when the system is near the LR end of the LR↔HR transition (Fig. 2b) where $K_{HR \rightarrow LR} \gg K_{LR \rightarrow HR}$ at $I_{ac} = 0$. According to Eq. (1), $K_{LR \rightarrow HR}$ is much more sensitive to small variations of current than $K_{HR \rightarrow LR}$ because $E_b^{HR} \ll E_b^{LR}$. Therefore, in the presence of $I_{ac}$, $K_{HR \rightarrow LR}$ is nearly time-independent while $K_{LR \rightarrow HR}$ oscillates with the frequency of the ac drive. This implies that at large enough $I_{ac}$, $K_{HR \rightarrow LR} \gg K_{LR \rightarrow HR}(t)$ for a fraction of the $I_{ac}$ period, while for another fraction of the period $K_{HR \rightarrow LR} \ll K_{LR \rightarrow HR}(t)$. If $T$ is high enough so that $K_{HR \rightarrow LR} \gg f$ and $\max\{K_{LR \rightarrow HR}(t)\}_t \gg f$, then the LR→HR→LR transition takes place in almost every cycle of $I_{ac}$. This results in large-amplitude resistance oscillations at low $f$ and gives rise to the large low-frequency tail in $V_{dc}(f)$. The negative sign of $V_{dc}(0)$ shows that the oscillations of $\vec{M}$ are in phase with the ac STT oscillations (more negative current favors the HR state), as expected for adiabatic SR. Figure 3f confirms the SR nature of the effect as the resonance turns on only at $T > T_{th}$, and quickly reaches the maximum amplitude at the SR temperature, $T_{SR}$. The turn-on of the SR is sharp in $T$ due to the exponential dependence of the transitions rates on $1/T$. The slow decay of $V_{dc}(0)$ for $T > T_{SR}$ is due to partial thermal randomization of the LR↔HR transitions[36].

The adiabatic SR is observed only for $I_{dc}$ at the LR end of the LR↔HR transition because the system returns to the LR state with current-sensitive $K_{LR \to HR}$ in almost every period of $I_{ac}$. In contrast, for $I_{dc}$ = -2.8 mA at the HR end of the LR↔HR transition, a low-frequency $I_{ac}$ does not induce the HR→LR transition because $K_{HR \to LR}$ is weakly sensitive to $I_{ac}$, and the system remains in the HR state. For this $I_{dc}$, only a signature of small-amplitude intra-well resonance in the HR state is expected in $V_{dc}(f)$. However, Fig. 2d shows that this is not the case. Although the low frequency tail in $V_{dc}(f)$ disappears at $I_{dc}$ = -2.8 mA, a peak due to unexpected high-frequency ($f$ = 2.2 GHz) large-amplitude dynamics is observed.

This surprising high-frequency dynamics can be explained if $I_{ac}$ excites large-amplitude oscillations of magnetization with a ~180° phase shift with respect to the ac STT drive. Fig. 4e illustrates how such a phase shift can result in a non-zero time-average component of STT perpendicular to the sample plane and thereby stabilize large-amplitude precession of $\vec{M}$ on an out-of-plane trajectory[40]. Since the angle between $\vec{M}$ and $\vec{M}_{Co}$ is not zero for H > 0, there is a non-zero component of STT perpendicular to the sample plane, $\tau_z \sim \vec{M} \times (\vec{M} \times \vec{M}_{Co})_z$. For $I_{dc}$ < 0, $\tau_z$ > 0 near the LR state and $\tau_z$ < 0 near the HR state. Therefore, if $\vec{M}$ oscillates on a large-amplitude trajectory passing near both the HR and LR states, the time-average $\tau_z$ due to $I_{dc}$ is close to zero. In contrast, $I_{ac}$ at the frequency of the oscillations of $\vec{M}$ on the large-amplitude trajectory generates positive time-average $\tau_z$ if the phase shift between the oscillations of $\vec{M}$ and ac STT is ~ 180°. Therefore, for motion of $\vec{M}$ phase locked to the ac STT with a ~ 180° phase shift, the STT from $I_{ac}$ always pushes $\vec{M}$ in the direction perpendicular to the sample plane (increases $M_z$), towards higher energy precessional trajectories. Figure 4c-d shows that for large enough $M_z$, $\vec{M}$ can precess on large-amplitude trajectories encircling both the LR and the HR energy minima. The frequency of precession on such high-energy trajectories can be lower than the frequency of the trajectories encircling only the HR or the LR energy minima[5]. Such a large-amplitude dynamic state (D) stabilized by $I_{ac}$ is consistent with the data in Fig. 2d showing a resonance with the amplitude corresponding to peak-to-peak resistance oscillations similar to $\Delta R$ and a frequency below the eigenmode frequencies in the LR and HR states. The width of the resonance peak in the D state is determined by the bandwidth of phase locking of the oscillations of $\vec{M}$ to $I_{ac}$ rather than by damping. This explains the large line width of the resonance in Fig. 2d, and the dependence of the line width on $I_{dc}$ in Fig. 3a. This also explains the increase of the line width with $I_{ac}$ shown in the inset of Fig. 3c because larger $I_{ac}$ increases the phase locking bandwidth[25]. The saturation of $V_{max}$ with $I_{ac}$ in Fig. 3c is due to the saturation of the amplitude of resistance oscillations in the D state at a value close to $\Delta R$ ($V_{max} \approx \frac{\sqrt{2} I_{dc} \Delta R}{2\pi}$).

Surprisingly, the amplitude of the D-state resonance in $V_{dc}(f)$ has almost the same temperature dependence as the adiabatic SR low frequency tail, $V_{dc}(0)$, (Fig. 3f). Therefore, the D state is thermally activated and the observed D-state resonance belongs to the class of SR phenomena. Since the frequency of the D state is greater than the Kramers transition rates between the HR and LR states, the D-state resonance is a high-frequency or non-adiabatic stochastic resonance (NSR)[41,42,43,44]. We now discuss the origin of the NSR effect in our SV system and explain why NSR is only seen near the HR end of the LR↔HR transition. To understand NSR, we consider thermally activated transitions among the HR, LR and D states. The time-average magnetic energies as well as the widths of the energy distributions in these states depend on STT as illustrated in Fig. 4b. Consideration of the available trajectories of $\vec{M}$ in the HR, LR and D states gives the relation between the time-average energies in these states: $\langle E_D \rangle > \langle E_{HR} \rangle > \langle E_{LR} \rangle$. At the HR end of the LR↔HR transition, where the NSR is observed, $K_{LR \rightarrow HR} \gg K_{HR \rightarrow LR}$, and thus if $\vec{M}$ falls from the D state to the low-energy LR state, it is rapidly returned to the higher-energy HR state as illustrated in Fig. 4f. Since $\langle E_{HR} \rangle$, is close to $\langle E_D \rangle$, small STT from $I_{ac}$ pushing $\vec{M}$ towards the D state is sufficient to supply energy $\langle E_D \rangle - \langle E_{HR} \rangle$ and thereby induce the HR→D transition in the half the period of $I_{ac}$ for which $\tau_z(t) > 0$. STT from $I_{ac}$ constantly supplies energy to the D state but not to the HR and LR states, for which ac STT adds energy for half a period of $I_{ac}$ and removes energy for the other half. Due to the energy supplied by $I_{ac}$ to the D state, this state becomes the most stable of the three non-equilibrium states (HR, LR and D) at sufficiently large $I_{ac}$ and $T$ resulting in $K_{HR \rightarrow D} \gg K_{D \rightarrow HR}$ and $K_{HR \rightarrow D} \gg K_{D \rightarrow LR}$. As a result, the system spends most of its time in the D state and a large-amplitude high-frequency peak appears in $V_{dc}(f)$ (Fig. 2d). Energy considerations also explain why NSR is not seen at the LR end of the LR↔HR transition. In the LR state, the energy gap to the D state is large $\langle E_D \rangle - \langle E_{LR} \rangle > \langle E_D \rangle - \langle E_{HR} \rangle$ and the energy supplied by ac STT is insufficient to induce the LR→D transition. A detailed quantitative understanding of NSR requires a Fokker-Planck description of the transition rates[23,45].

The large rectified voltage due to NSR can be employed for microwave signal detection[46]. Indeed, by using high-magnetoresistance MgO magnetic tunnel junctions (MTJ)[47,48], a rectified voltage of greater than 0.1 V in response to microwave currents of ~ 100 μA may be expected at NSR. The expected sensitivity of such an MTJ-based detector, ~20 mV/μW, is greater than that of Schottky diode detectors.

In conclusion, we observe adiabatic and non-adiabatic SR of magnetization co-excited by ac STT and temperature in nanoscale spin valves. Our work demonstrates that combined dc and ac STT applied to a nanomagnet stabilize unusual dynamic states of magnetization far from equilibrium. The rates of thermally activated transitions from these states are sensitive to the amplitude and phase of the applied STT, and thus nanomagnets driven by ac and dc STT[24,25] provide a convenient playground for studies of thermodynamic processes far from equilibrium. The amplitude of magnetization oscillations in the non-adiabatic SR regime is nearly two orders of magnitude greater than that in the FMR regime for the same level of the ac STT drive. The non-adiabatic SR dynamics give rise to large rectified voltages generated by spin valves in response to applied microwave currents, and thus non-adiabatic SR can be utilized in sensitive microwave signal detectors of nanoscale dimensions.

X. Cheng et al.                                      9                             December 18, 2009

**Acknowledgements**

This work was supported by the NSF (grants DMR-0748810 and ECCS-0701458) and by the Nanoelectronics Research Initiative through the Western Institute of Nanoelectronics. We thank D. Ralph and R. Buhrman, in whose labs the samples were made, for helpful discussions and assistance with the sample preparation. We acknowledge the Cornell Nanofabrication Facility/NNIN and the Cornell Center for Nanoscale Systems, both supported by the NSF, which facilities were used for the sample fabrication.

**Author contributions**

X.C. collected and analyzed data and wrote the paper; I.K. made the samples and wrote the paper. All authors contributed to the data collection and the preparation of the manuscript.

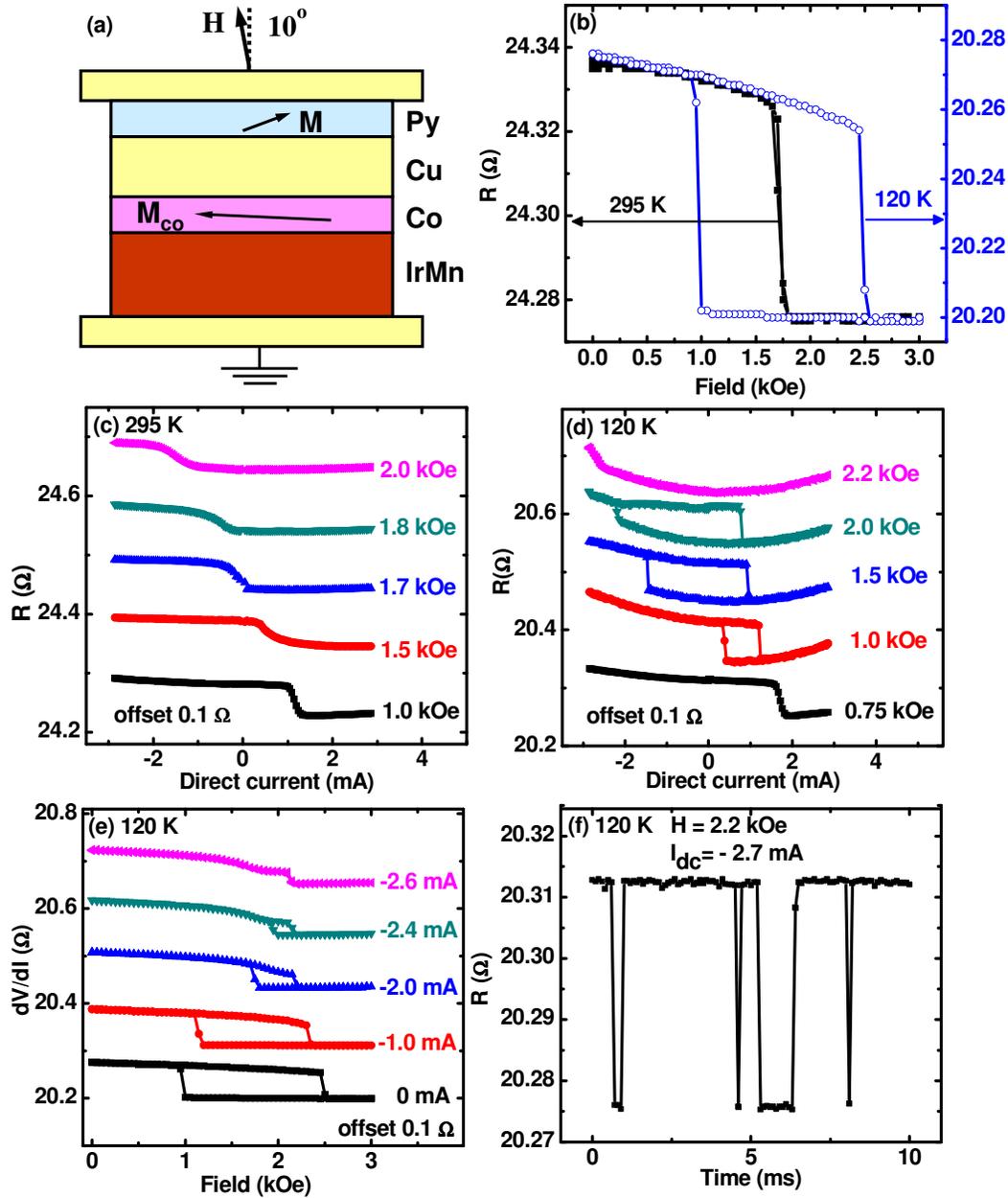

**Figure 1| Spin valve magneto-resistance and current-induced switching.** **a**, Schematic view of the spin valve with approximate directions of magnetization of the pinned, $\vec{M}_{Co}$, and the free, $\vec{M}$, layers in the high resistance (HR) state of the spin valve. External magnetic field, $\vec{H}$, is applied at 10° to the sample plane normal. **b**, Resistance $R$ as a function of $H$ at two temperatures: 295 K and 120 K. $R$ as a function of direct current, $I_{dc}$, at 295 K (**c**) and 120 K (**d**). **e**, Differential resistance, $dV/dI$, as a function of $H$ at 120 K measured for several values of $I_{dc}$. **f**, Telegraph noise of $R$ between the high (HR) and low (LR) resistance states at $T$ = 120 K, $H$ = 2.2 kOe and $I_{dc}$ = -2.7 mA.

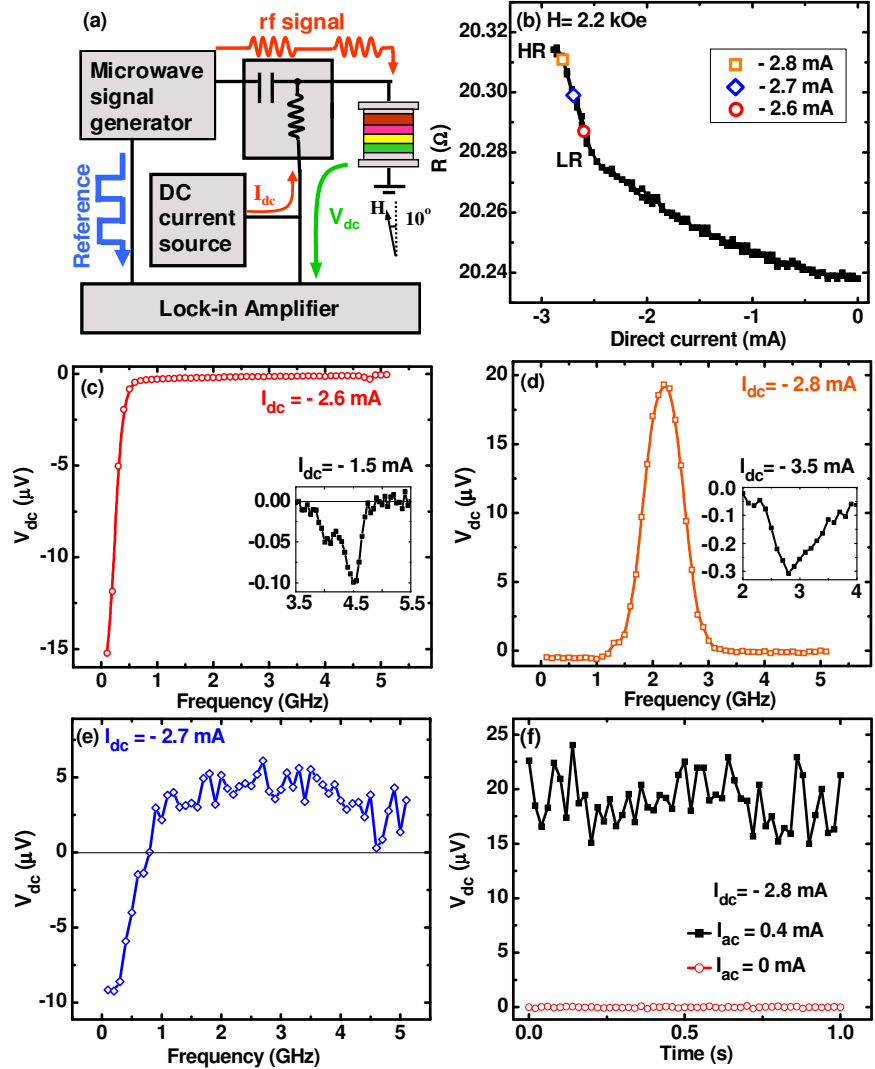

**Figure 2| Measurements of magnetization dynamics driven by ac spin transfer torque. a**, Spin torque FMR (st-FMR) measurement setup. **b**, $R$ versus $I_{dc}$ at $T = 120$ K, $H = 2.2$ kOe. Symbols mark the direct current values for which st-FMR data are shown in **c**-**e**. Measured response curves, $V_{dc}(f)$, at the LR↔HR transition: **c,** $I_{dc}$ = -2.6 mA, **d,** $I_{dc}$ = -2.8 mA, **e,** $I_{dc}$ = -2.7 mA. Insets: st-FMR spectra far from the LR↔HR transition (**c,** $I_{dc}$ = -1.5 mA) and (**d** , $I_{dc}$ = -3.5 mA). **f**, $V_{dc}(t)$ at $I_{dc}$ = -2.8 mA, for $I_{ac}$= 0 (open circles) and $I_{ac}$= 0.4 mA at $f$ = 2.2 GHz (solid squares). The data in **c**-**e** are measured at $T$ = 120 K, $H$ = 2.2 kOe, $I_{ac}$= 0.4 mA.

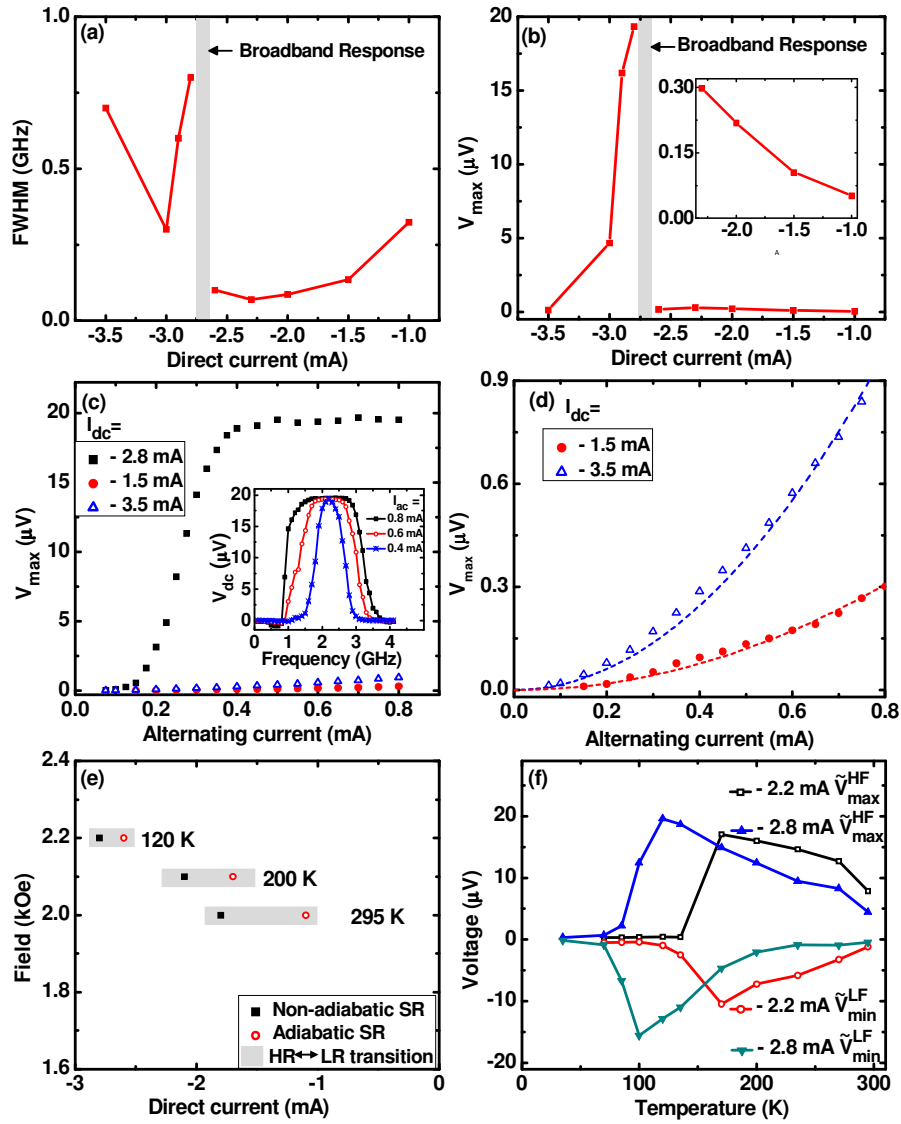

**Figure 3| Dependence of ac-STT-driven magnetization dynamics on current and temperature. a,** Full width at half maximum and **b,** the amplitude, $V_{max}$, of the high-frequency peak in $V_{dc}(f)$ response curves as functions of $I_{dc}$. Grey bands in (a) and (b) mark the crossover regime between the large-amplitude high- and low-frequency peaks. In this crossover regime, broadband response curves $V_{dc}(f)$ such as that shown in Fig 2(e) are observed. Inset in **b**: blow-up of the low-current region of $V_{max}(I_{dc})$. **c,** The dependence of $V_{max}$ on ac drive current, $I_{ac}$, for three values of $I_{dc}$: far from (-1.5 mA, -3.5 mA) and at (-2.8 mA) the LR↔HR transition. Inset: $V_{dc}(f)$ response curves in the regime of large $I_{ac}$ for $I_{dc}$=-2.8 mA. **d,** Quadratic dependence of $V_{max}$ on $I_{ac}$ for $I_{dc}$ far from the LR↔HR transition. Lines are quadratic fits to the data. **e,** Phase diagram of the system in the $(H, I_{dc})$ plane. Grey bands mark regions of the LR↔HR resistance transitions at fixed field and temperature. Solid squares and open circles mark the direct current values at which maximum rectified signal due to adiabatic and non-adiabatic SR are observed at fixed field and temperature. **f,** Temperature dependence of the amplitudes of the high frequency $\tilde{V}_{max}^{HF}$ and low frequency $\tilde{V}_{min}^{LF}$ st-FMR signals at two values of $I_{dc}$: -2.2 mA (open symbols) and -2.8 mA (solid symbols).

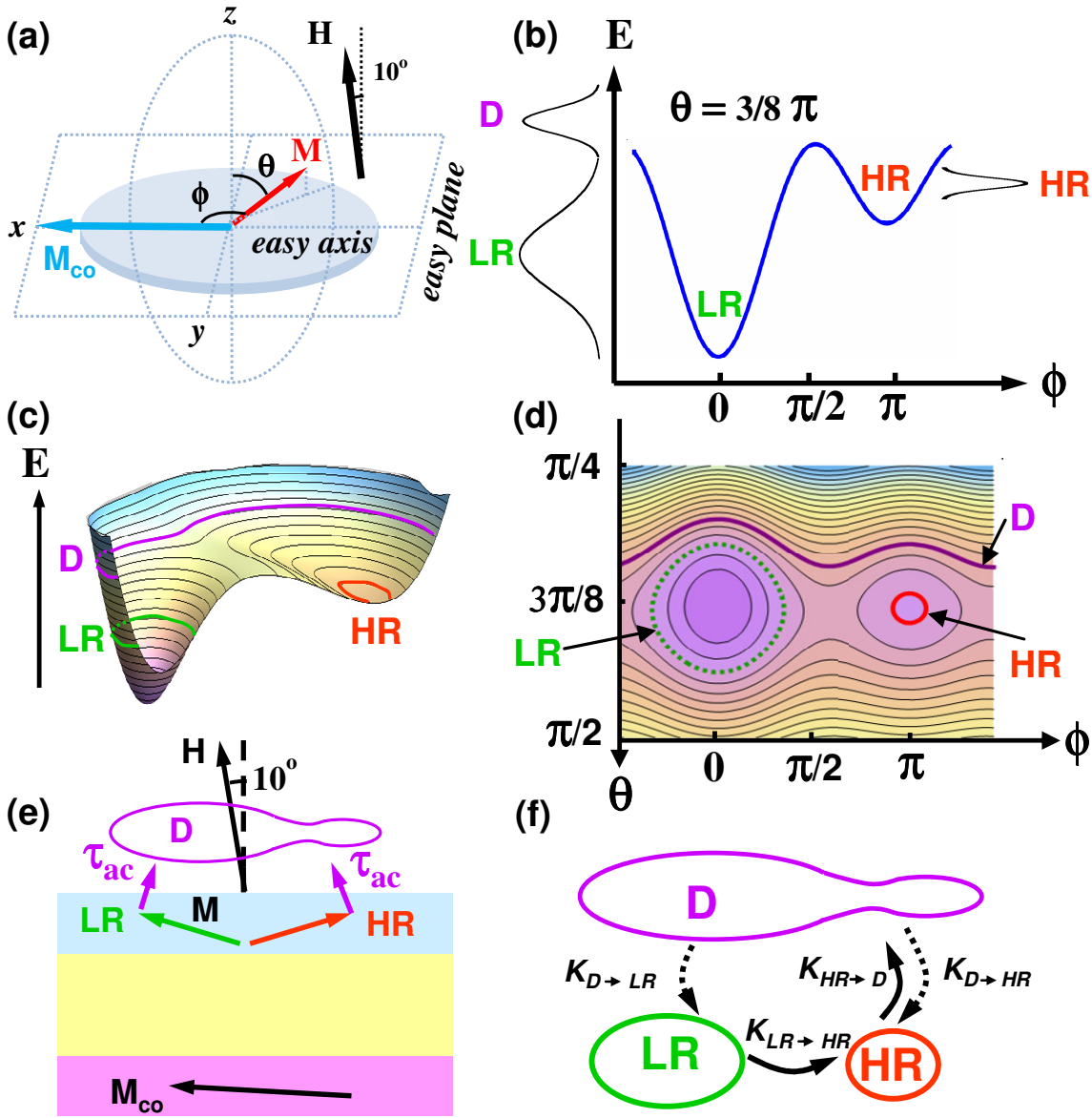

**Figure 4| Stochastic resonance energy diagrams. a,** Spherical coordinate system in which the energy of the Py nanomagnet, $E(\theta,\varphi)$, is described by Eq. (2). **b,** $E(\theta,\varphi)$ as a function of $\varphi$ for $\theta = 3\pi/8$, the approximate equilibrium polar angle of the Py magnetization in the HR and LR states. This energy is a double-well potential with the minima corresponding to the HR and LR states. Schematic energy distributions of magnetization in the HR, LR and the dynamic state D at non-zero $I_{dc}$ and $I_{ac}$ are shown. **c,** 3D sketch of the energy surface $E(\theta,\varphi)$; thick solid lines schematically show magnetization trajectories with time-average energies, $\langle E_{LR}\rangle$, $\langle E_{HR}\rangle$, $\langle E_D\rangle$ in the LR, HR and D states. **d,** Contour plot of $E(\theta,\varphi)$; dashed lines schematically show $\langle E_{LR}\rangle$, $\langle E_{HR}\rangle$, $\langle E_D\rangle$. **e,** Alternating STT, $\tau_{ac}$, always pushes magnetization out of the sample plane towards high energy trajectories of the D state if the phase of the magnetization oscillations is ~ 180° with respect to the ac STT drive. This $\tau_{ac}$ stabilizes the large-amplitude dynamic state D. **f,** Sketch of the Kramers transition rates among the HR, LR and D states at the non-adiabatic stochastic resonance. Solid arrows show the dominant path bringing magnetization into the D state. Dashed arrows show transitions out of the D state.